\documentstyle[referee]{mn}
\newcommand{\reference}{\bibitem}
\def\beq{\begin{equation}}
\def\eeq{\end{equation}}
\def\bey{\begin{eqnarray}}
\def\eey{\end{eqnarray}}
\def\beqarray{\begin{eqnarray}}
\def\eeqarray{\end{eqnarray}}
\def\apj{ApJ}

\def\rs{r_{\rm s}}

\def\zs{z_{\rm s}}
\def\zc{\bar{z}}
\input epsf
\title[]        
{Predicting the Second Caustic Crossing in Binary Microlensing Events}
\author[Jaroszy\'nski \& Mao]
{
Micha{\l} Jaroszy{\'n}ski$^{1,3}$, Shude Mao$^{2,3}$\thanks{e-mails:
mj@astrouw.edu.pl, smao@jb.man.ac.uk}\\
$^1$Warsaw University Observatory, Al. Ujazdowskie 4, 00-478 Warsaw, Poland\\
$^2$University of Manchester, Jodrell Bank Observatory,
  Macclesfield, Cheshire SK11 9DL, UK \\
$^3$Princeton University Observatory, Princeton, NJ 08544-1001, USA}
\date{Accepted ........
      Received .......;
      in original form .......}
\pubyear{2001}

\begin{document}
\maketitle
\begin{abstract}
We fit binary lens models to the data covering the initial part of real 
microlensing events in an attempt to predict the time of the second 
caustic crossing. We use approximations during the initial search 
through the parameter space for light curves that roughly match the
observed ones. Exact methods for calculating the lens magnification
of an extended source are used when we refine our best initial models. 
Our calculations show that the reliable prediction of the second crossing
can only be made very late, when the light curve has risen appreciably 
after the minimum between the two caustic-crossings. The best
observational strategy is therefore to sample as frequently as possible
once the light curve starts to rise after the minimum.
\end{abstract}


\section{INTRODUCTION}

Since Paczy\'nski (1986) first proposed to use microlensing
as a method of detecting compact dark matter objects in the Galactic
halo, the field has made enormous progress (see e.g., Paczy\'nski
1996; Mao 1999 for reviews). Several microlensing searches have yielded
more than one thousand microlensing events and many more variable stars 
(e.g. Alcock et al. 2000c; Alcock et al. 2000d; Beaulieu \& Marquette 2000;
Szyma\'nski et al. 2000; Udalski et al.  2000). The two dozen microlensing
events toward the Large Magellanic Clouds indicate that compact halo 
objects do not make up 100\% of the halo (e.g. Lasserre et al. 2000;
Alcock et al. 2000e). The microlensing technique turns out to
be extremely useful for a variety of other purposes, such as studying
mass-functions, Galactic structure and stellar atmospheres. Many
of the microlensing conclusions (including the fraction of mass in
compact form) are subject to small number statistics: e.g., the published 
optical depth toward the Large Magellanic Cloud are based on $\sim 15$
events (Alcock et al. 2000d), while that toward the bulge is based on 
$\sim 100$ events (Alcock et al. 2000b; see also Alcock et al. 1997;
Udalski et al. 1994c). This is because although the number of light 
curves has reached one thousand (e.g. Wo\'zniak 2000), the detection 
efficiency curve has not been yet calculated,
so the usefulness of these events is somewhat limited; this difficulty
may soon be removed, however (Wo\'zniak 2001, in preparation). 

About 10\% of the observed
 microlensing events are  binary events (e.g. Udalski et al. 2000,
Alcock et al. 2000a), as predicted by Mao \& Paczy\'nski (1991). The
caustic crossing binary events among these are extremely important for
several reasons. First, their binary signature is unique (see section 4 for
examples), and hence they are easily recognizable. Second, the caustic 
crossing induces very high magnifications, and therefore they are useful for
intense photometric and spectroscopic follow-ups; such observations can
be used to study stellar atmospheres, probe the age,  and metallicity of
main sequence stars in the Galactic  bulge (e.g. Albrow et al. 1999a;
Lennon et al. 1996, 1997; Sahu \& Sahu 1998; Minitti et al. 1998;
Heyrovski, Sasselov \& Loeb 2000). 
Thirdly, the caustic crossings always come in pairs,
so once we observe the first crossing, if we can predict the second 
caustic crossing, then we can time our observations more accurately. 
A question naturally arises: is it possible to predict the
second-caustic crossing based on the data prior to it?
This is a timely question
because the OGLE collaboration is currently upgrading
their instruments from OGLE II to OGLE III. Once finished, OGLE III 
will discover hundreds of microlensing events each year, perhaps 5\% 
of these will be caustic-crossing binary 
microlensing events. The primary motivation of this paper is to address this
question. 

The layout of the paper is as follows. In section 2 we first give the
lens equation, and outline our numerical methods. In section 3 we
describe the algorithm of searching the binary lens parameter space.
And in section 4 we apply our methods to two real-time binary events 
discovered by the OGLE II collaboration. In section 5, we discuss several
issues in fitting binary lenses.

\section{NUMERICAL METHOD}

\subsection{Lens equation}

We use the complex notation  of the binary lens equation
(Witt 1990)
\beq \label{eq:lens}
\zs = z - {m_1 \over {\zc-z_1}}
- {m_2 \over {\zc-z_2}} = z- {\zc \over (\zc - z_1) (\zc-z_2)}~,
\eeq
where $\zs$ is the source position, and $z_1$ and $z_2$ are the two lens
positions. The total mass of the two lenses is normalized to one 
($m_1+m_2=1$), and we choose the coordinate system such that 
the lenses are on the $x$-axis, and the ray crossing the origin passes
undeflected, so $m_1 z_2 + m_2 z_1 = 0$; the second step in 
eq.~(\ref{eq:lens}) follows from our choice of the coordinate system.

\subsection{Light curve for point sources}

The lens equation can be manipulated into a fifth order polynomial by
taking the conjugate of eq.~(\ref{eq:lens}) and substituting the
expression of $\zc$ back into the lens equation. 
 The associated polynomial can be readily solved using
well-known numerical schemes (e.g., Press et al. 1992); the image
positions and magnifications can be found
for any source position. The efficiency can
be further improved by combining the brute-force polynomial solver and
Newton-Ralphson method. Essentially, we use the image positions
found from the previous step as the initial guess solutions for the new
source position. Usually, this allows quick
convergence to either three or five solutions. These solutions
are then deflated from the polynomial, which results in a lower order
(usually quadratic) polynomial that can be readily solved.
We find that this method speeds up the finding of the image locations
by at least a factor of few, depending on the machine architectures.

\subsection{Light curve for extended sources}

The magnification of an extended source with arbitrary
surface brightness profiles can in principle be found by a two-dimensional
integration over the stellar surface. However, this is in general very
time-consuming. For axis-symmetric sources, considerable speedup can be
achieved using the Stokes theorem. In this case, we only need to solve
the lens equation for points belonging to the boundary circle (Gould \&
Gaucherel 1997; Dominik 1998). The magnification is obtained by
appropriate weighting of  the magnification of points on the circle
according to the source profile (see Dominik 1998). The details of our 
implementation can be found in Mao \& Loeb (2001).

\section{FITTING BINARY LIGHT CURVES}

Many authors have given examples of binary lenses models which fit
particular cases of events with light curves showing the characteristics
of caustics crossing. The important property of all these models is 
their non-uniqueness, especially in cases when the coverage of the 
caustic crossing part of the light curve is missing or weak
(Mao \& Di Stefano 1995). A good 
example is the first observed binary lens event OGLE \#7 (Udalski et 
al. 1994b), which can be fitted with several different models as 
shown by Dominik (1999). Even in cases with very dense observations 
including caustic crossing the fit may be not unique, as shown by 
Afonso et al. (2000) for the event MACHO 98-SMC-1.

The prediction of the second caustic crossing is even a less constrained
problem. The observations of the first crossing are usually sparse and
the data containing most useful information are still missing. There must
be many models of the event giving equally good fits to the data. It is
not excluded, however, that the time of the second caustic
crossing can be estimated.

In the numerical experiments we use only the data representing
the early parts of the light curves.  We also change the amount of data
including observations made on subsequent nights and check how it
influences our predictions.

We assume that the data already acquired shows the characteristics
of caustic crossing event, i.e. a strong increase of brightness followed
by a slower decline resembling the beginning of the typical ``U-shaped''
light curve.  We also assume that
the inter-caustic minimum of brightness is already covered by the data. 
If this is the case one can estimate the total brightness (energy flux) 
in three characteristic instants of time: long before the event (``base
flux'' $F_0$), shortly before the first caustic crossing ($F_1$) and at 
the inter-caustic minimum ($F_{12}$).  The time, $t_{12}$, corresponding
to the flux minimum can also be estimated.  The base flux $F_0$ is usually
measured with hundreds of data points, and we neglect its  error. The other
fluxes  are estimated with errors which we take into account.
Despite the lack of accuracy the estimates may be used to reject some
lens models thus diminishing the volume of the possible parameter space.
The measured flux comes from the source of interest, other stars within
the telescope beam, and possibly from the lens. The source contributes
some fraction $f$ ($f \le 1$) of the base flux, and only this part ($fF_0$)
is amplified by the lens. The remaining part $(1-f)F_0$ is not
changing.  The observed fluxes are related by the equations:
\beq \label{eq:flux-before}
F_1 = (fA_1 + 1 -f)F_0~,
\eeq
\beq \label{eq:flux-inside}
F_{12} = (fA_{12} + 1 -f)F_0~,
\eeq
where $A_1$ and $A_{12}$ are the lens magnifications corresponding to
$F_1$ and $F_{12}$. Since the source contributes less than 100\% of the
flux ($f \le 1$), the following conditions apply:
\beq \label{eq:A1-ineq}
A_1 \ge \min\left[{F_1 \over F_0}\right]~,
\eeq
\beq \label{eq:A12-ineq}
A_{12} \ge \min\left[{F_{12} \over F_0}\right]~.
\eeq
One can also obtain the following inequalities:
\beq \label{eq:combined-ineq}
\min\left[{F_{12}-F_0 \over F_1-F_0}\right]
\le {A_{12}-1 \over A_1-1}
\le \max\left[{F_{12}-F_0 \over F_1-F_0}\right]~,
\eeq
where $\min$ / $\max$ stand for the minimal / maximal values allowed by
the estimates, including their 3-$\sigma$ errors.

We assume that the observations made during the first caustic crossing
are sparse and so insufficient to find the size of the source and its
limb darkening profile. The only signature of the caustic crossing is
the implied discontinuity in the light curve and the presence of at
least one point with high magnification. The approximate description of
caustic crossing by a small source (Witt 1990) may be used to obtain 
an upper limit on its radius $\rs$. According to this approach the lens
magnification has a generic form near the caustic, and the observed flux
can be expressed as:
\beq \label{eq:close}
F = \left(fA_1+f{KG(s_\perp/\rs) \over \sqrt{\rs}}+1-f\right)F_0~,
\eeq
where $K$ is a constant depending solely on the caustic properties at
the crossing point, which can be calculated using the prescription of
Witt (1990). The shape of function $G$ depends weakly on the limb
darkening profile; we have to neglect this factor using the uniform disk
(no limb darkening) as a source model. The distance $s_\perp$ is
measured along an axis perpendicular to the caustic at the crossing
point and directed inward. 

The maximum magnification during the caustic crossing corresponds to the
maximum of function $G$. The maximal measured flux ($F_{\rm max}$) 
cannot exceed the value allowed theoretically. Transforming the above 
equation into inequality and using 
eqs.~(\ref{eq:flux-before}-\ref{eq:flux-inside}) 
to substitute for $f$ we get:
\beq
\sqrt{r_{\rm s(max)}} \le 
{F_{12}-F_1 \over F_{\rm max}-F_1}{KG_{\rm max} \over A_{12}-A_1}~,
\eeq
where all quantities in the RHS are either measured or can be calculated
from the model being fitted. Due to the errors in measured fluxes the
upper limit on the source size is only a rough estimate.

The probability, that the observed maximal flux actually corresponds to
the maximal lens magnification, is essentially zero. Sources larger than
the limit never reach the measured maximal brightness and are excluded.
For sources smaller than the limit, the brightness remains higher than 
the measured maximal flux $F_{\rm max}$ for a finite time, so the 
probability of measuring the flux of at least this
value is positive.  The (unobserved) part of the light curve, when the
flux remains higher than the highest flux observed,  has the longest 
duration for sources about two times smaller than the estimated maximal
size $r_{\rm s(max)}$. For even smaller sources the duration of this phase 
is slightly shorter; it goes to zero only for sources approaching the 
maximal size. 

Except for the introduced upper limit the size of the source cannot be
constrained further. We use $\rs=0.5 r_{\rm s(max)}$ as a
likely but still ad hoc choice.

\subsection{Monte Carlo Simulations}

We limit ourselves to static binary lenses. Since we can put only weak
limit on source size, we neglect limb darkening. With
these simplifications, we are left
with seven unknown parameters (mass ratio $q \equiv m_1/m_2$, binary
separation $d$ expressed in units of Einstein radius,
direction of the source motion given by the angle $\beta$ between its
trajectory and the line joining the binary members, 
source encounter parameter $b$ relative to the origin of the coordinate system,
times of the first $t_1$ and second $t_2$ caustic crossing and the 
parameter defining the source contribution to the base flux $f$). Since 
once other parameters are fixed the best  $f$ can be found analytically
(see below), the parameter space which has to be investigated numerically
has six dimensions.

We need a time-efficient scheme to look for the solutions in the multi
dimensional space. There is a natural hierarchy of the parameters, which
we follow. The most important are the physical parameters of the binary
($q$, $d$) which also define the caustic structure of the model. Given
the caustic pattern and the source path direction ($\beta$) we can
find the range of possible encounter parameters ($b$) leading to caustic
crossings. For given source trajectory the caustic crossings are located
at some positions  $s_1$, $s_2$ along the path, and the minimum of lens 
magnification corresponds to position $s_{12}$.  The lens magnification as
a function of the source position can be found from the model.  Knowledge of
the crossing times $t_1$, $t_2$ is only needed to translate it into the 
usual time dependence of the light curve.

We start from choosing the binary parameters ($q$, $d$), and find the
corresponding caustic pattern. We use a grid with spacings 
$(\Delta q, \Delta d)=(0.02, 0.02)$ for close binaries,  and 
$(\Delta q, \Delta d)=(0.04, 0.05)$ for intermediate or wide systems
(compare Afonso et al. 2000). Our search spans the full range of mass
ratio ($q \in (0.02, 1)$) and a wide range of separations ($d \in (0.1,
10.)$). 
The source direction ($\beta$) is searched on a grid with
$\Delta\beta=3^\circ$. 
For each $\beta$ we find the range of possible encounter parameter
values. We search the allowed range of $b$ using Monte Carlo method. 
For intermediate binaries the caustic pattern consists of a single closed
curve and we assume equal probability for all possible values of
encounter parameter. For close or wide systems the caustic pattern consists of
three or two disjoint closed curves and there may be more than
one separate ranges of $b$. If this is the case, we make the same number
of Monte Carlo shots for encounters with each of the caustic curves, and
assume uniform probability distribution for $b$ within the range corresponding
to any of them.

During the extensive search through the parameter space we use
approximations to speed up the calculations (Compare Albrow et al. 1999b).
We use point source
magnification everywhere, except in the close vicinity of caustics, where we
use generic light curve shape of eq.~(\ref{eq:close}). The magnification
is calculated for many points along the source path and stored 
(as function of the position $A(s)$) for further use.
The time of the first crossing $t_1$ can be bracketed by an analysis of the
observations. The analysis yields also the time of the inter-caustic
minimum $t_{12}$ and its error. Using Monte Carlo we choose $t_1$ and
$t_{12}$ from their allowed ranges. (It is equivalent to setting $t_1$
and $t_2$). Using the correspondence between time and source position we find
the lens magnifications at the time of observations $A_i=A(s(t_i))$ by
interpolation. Now we estimate the goodness of fit:
\beq
\chi^2 = \sum_{i=1}^N {\left((fA_i+1-f)F_0-F_i\right)^2 \over \sigma_i^2}~,
\eeq
where $N$ is the number of observations, $F_i$ are the measured fluxes, 
and $\sigma_i$ - their errors. We use the value of parameter $f$ from the 
condition  $\partial\chi^2/\partial f = 0$, which is a linear equation.

We store parameter values for all the models for which the calculated
approximate $\chi^2$ has a value smaller than $2$ per degree of freedom.
Also the best models for given binary mass ratio and separation $(q,d)$ 
are stored. We check these models
repeating the $\chi^2$ calculation using extended sources and no
interpolation. Finally we refine our calculation for $\approx 100$
models with lowest $\chi^2$ allowing for small variations in all
parameters, which we choose again using Monte Carlo method.
Whenever we find a model with lower $\chi^2$ we treat it as a temporary 
solution and look for further improvement by varying its parameters.

\section{APPLICATIONS}

We apply our method to the two events observed by the OGLE II Experiment
(Udalski, Kubiak, \& Szyma\'nski, 1997)
dubbed 2000-BUL-38 and 2000-BUL-46, which were discovered by the 
Early Warning System (Udalski et al. 1994a).

Now, when the events are over, we
check predictions that would be made one to six days before the
actual second crossings of caustics. We simulate such predictions trying
to fit binary lens models to the observed light curves using the incomplete
data sets corresponding to the early part of observations.

\subsection{2000-BUL-38}

We have checked the variability of this source in previous observational
seasons.  While the visual inspection of the light curve may suggest
some kind of quasi-regular variability, we have not been able to find 
any periodicity in the data. We have also checked the hypothesis that
the source had a constant brightness before the event of 2000 season.
The averaged base flux of the source $F_0$ corresponds to $I$-band
magnitude $I_0=17.87$,  in agreement with public domain data of the 
OGLE II collaboration (Udalski, Kubiak \& Szyma\'nski 1997). 
We also use the photometric errors estimate from OGLE II database.
We apply the $\chi^2$ test to the model assuming constant flux of the
source. The test gives  the $\chi^2$ value much higher than acceptable
limits. We have to multiply all the errors by $1.73$ to get the $\chi^2$
value of $\approx 1$ per degree of freedom. In further analysis we apply such
adjustment also to the errors of the season 2000.

\begin{figure} 
\epsfxsize=\hsize
\begin{center}
\epsfbox{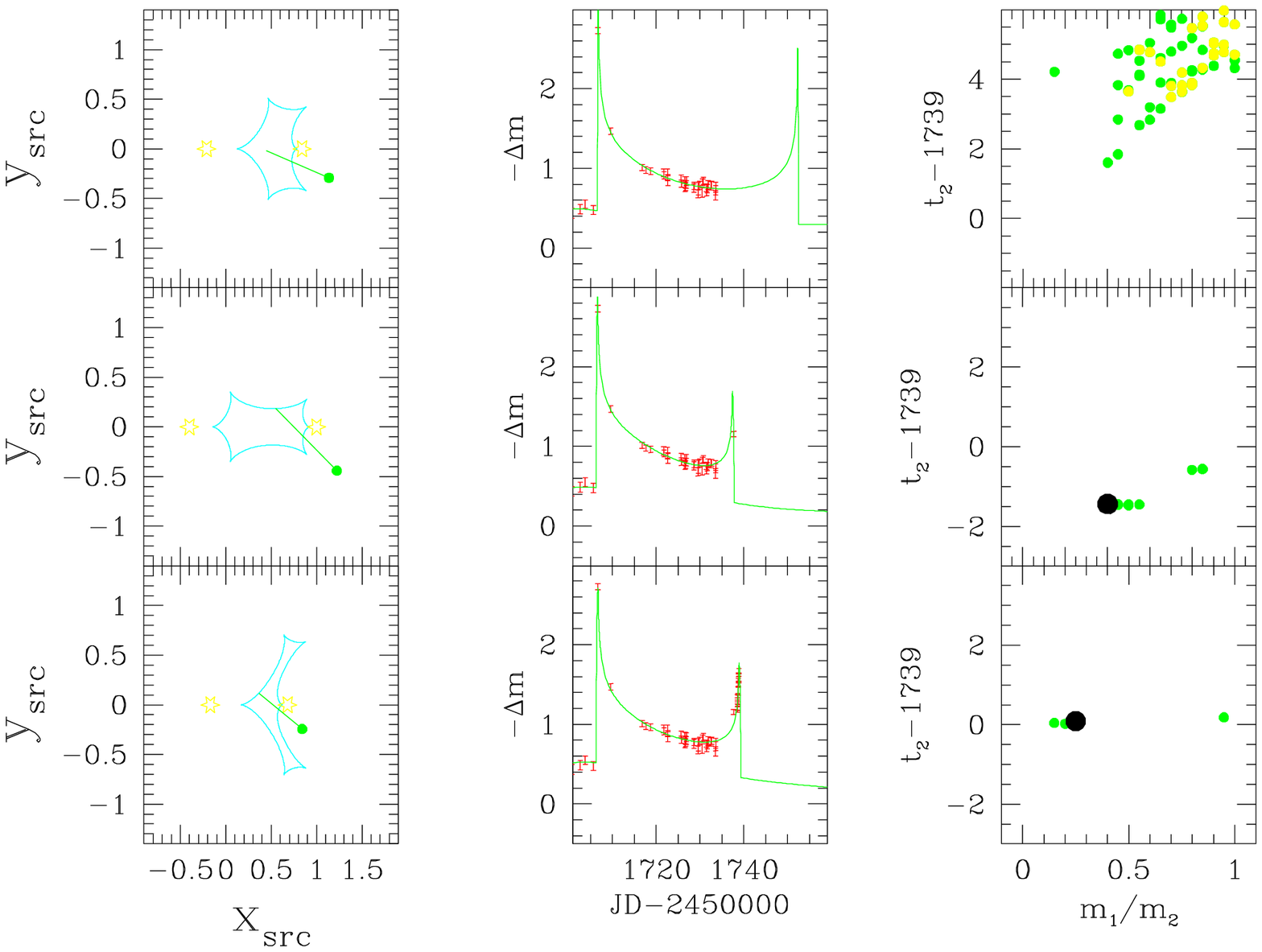}
\end{center}
\caption{Predictions for the event 2000-BUL-38. In the upper row we show
the results based on the data for JD$<2451734$ - six nights before the actual
second caustic crossing. The left panel is based on our best fit and 
shows the source plane with caustic pattern, the source path
corresponding to the used data and the projected position of the 
binary (stars).  The middle panel shows the corresponding theoretical 
(solid line) light curve and the data  (error bars ).
In the right panel we show the predicted time of the
second caustic crossing for the best fit (big dot - in this case out of
the range) and other acceptable models  (small dots). 
Similar results based on the data for $JD<2451738$ and
$JD<2451739$ are shown in the middle and lower row respectively.}\label{b38}

\end{figure}

For the model fitting we discard the data from previous seasons as
already used in the $F_0$ estimate. We use only measurements starting
from the point, when magnification exceeds $A=1.3$, since the details of
the binary lens model have little influence on the low magnification
tails of the light curve. We make numerical experiments for 2000-BUL-38 using
data accumulated before: $JD-2450000=1734$, $1738$, or $1739$. (These Julian
dates correspond to nights six, two, and one day before the actual second
caustic crossing). The first data sample constrains the models very
weakly. Various intermediate separation binaries with full range of mass
ratios as well as some close binaries with low mass ratio give
acceptable fits to the observations. The second caustic crossing time 
$t_2$ can not be predicted: the value of this parameter corresponding 
to different
acceptable models has a large spread; all acceptable models give too
late $t_2$. The geometry of the source - binary encounter, the light
curve for the best model and the predicted time of the second crossing
based on the first considered data sample are shown in the upper row of
panels in Fig.~\ref{b38}.

The second data sample has only one extra point -- the single
observation made after three nights of no data. All acceptable models
have intermediate separation  $d \approx 1.5$ and mass ratio $q \ge
0.4$. The predicted second crossing time is either one or two nights too
early. (Compare the middle row in Fig.~\ref{b38}.)

Finally we use the data sample including the night preceding the second
caustic crossing. The best model is marginally consistent with the data.
Even the models with much higher $\chi^2$ all predict the time of the
second crossing correctly. (See Fig.~\ref{b38}.)

The inspection of Fig.~\ref{b38} shows that the ``best'' models chosen by
our procedures based on different amount of data are not the same. One
may think that the accumulation of the data can only serve as to reject
some of the models, so the latest fits should be present among the
earlier results. To clarify this point we take $\approx 10^2$ of our
lowest $\chi^2$ fits obtained on the latest day considered, remove the
observations of the previous night and refine the fits allowing for
small changes in the encounter geometry, source size, and timing. The
parameter which changes most appreciably during the refinement is the
time of the second crossing; for all models considered its value
decreases and becomes $\approx 1$ day too early. The single observation
of $JD=2451737.67$ remains on the growing part of the light curve. (The
jump to the other side of the caustic is not excluded by the refinement
procedure, but the placing of the observation point on steeply falling
part of the theoretical light curve has low probability.) 
The models selected by the fits to the $JD < 2451939$ data sample and
then fitted to $JD < 2451738$ data sample remain (after refinement)
significantly  worse than the models selected from the beginning with 
the fits to $JD < 2451738$ data sample. This shows that the model
preferred by the data may change when the data is accumulated.

\subsection{2000-BUL-46}

\begin{figure} 
\epsfxsize=\hsize
\begin{center}
\epsfbox{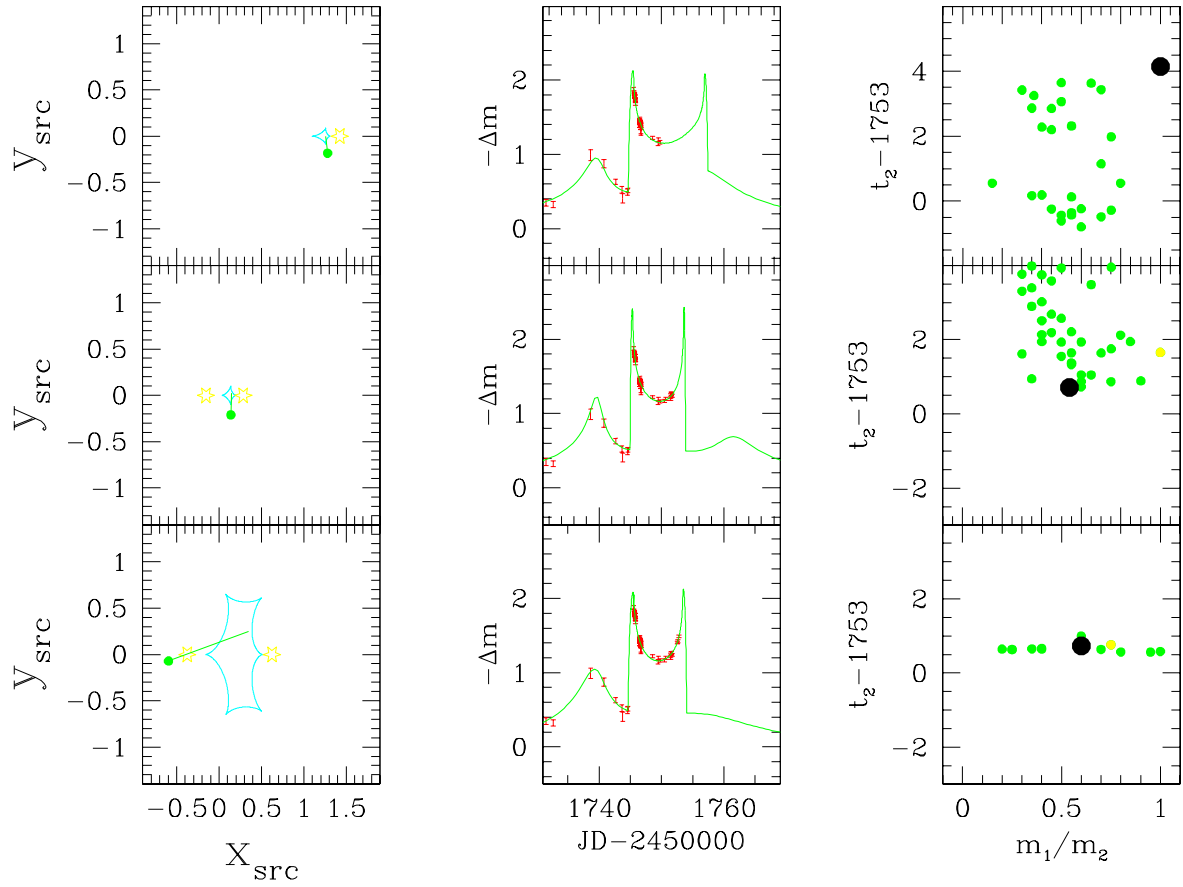}
\end{center}
\caption{Predictions for the event 2000-BUL-46. The conventions follow
Fig.\ref{b38}. The results correspond to the data for $JD<2451750$
(upper row), $JD<2451752$ (middle), and $JD<2451753$ (lower), which are 
respectively four, two, and one night before the second 
crossing.}\label{fig:b46}
\end{figure}

We treat this event with a manner similar to 2000-BUL-38. We adjust
the errors using a much lower factor ($1.11$). We consider three data
samples accumulated before $JD-2450000=1750$, $1752$, or $1753$, which
correspond to observations made four, two, or one day before the
caustic crossing. Because the event begins with the source passing
close to the cusp and then through the caustic, the models are better
constrained than for the other event considered. 
The first data sample can be fitted by models representing the three
families of binary lenses (close, intermediate, and wide).  The fits are
marginally consistent with the data. The best model is a wide binary. 
Predicted crossing times are from one day too early to 
four days too late. (Compare Fig.\ref{fig:b46}.) 
The marginal fits to the second data sample are similarly distributed in
the plane of physical binary parameters $(q,d)$. 
A close binary with the correct prediction of the
crossing gives the best fit now, but other marginally consistent models
give crossings up to four days too late.
The last data sample leaves us only with intermediate and wide binaries, 
and the best model belongs to the former class. All the marginally
consistent models give the right crossing time within accuracy of
$\pm 0.2\mathrm{d}$, except one, for which the prediction is
$\approx 0.5\mathrm{d}$ too late.

\subsection{Fitting ``square root'' formula}

For comparison we use a method which can only be applicable to the 
part of the data representing flux increase toward the caustic. For a
point source close to the crossing one can use a generic form of the
light curve given by the formula:
\beq \label{eq:sqrt}
F(t)=F_2+{\tilde{K} \over \sqrt{t_{\rm pred}-t}}~.
\eeq
There are three unknown parameters: the flux measured shortly after the 
crossing $F_2$, a constant $\tilde{K}$, which is proportional to $K$ of 
eq.~(\ref{eq:close}), and the time of the crossing $t_{\rm pred}$. 
Having three exact measurements of the flux
in the region of formula validity, one would be able to calculate the time
of the crossing. For any three points representing monotonically
increasing flux, such that the middle point is below the straight line
joining the other two points, the fit of the above formula is possible
and gives some $t_{\rm pred}$. Since the actual light curve is not well
approximated by the formula far from the crossing or very close to the
crossing, when the limited size of the source becomes important, one
expects quite large scatter in fitted values of $t_{\rm pred}$ . We
simulate the process of applying such a procedure. 

\begin{figure} 
\epsfxsize=\hsize
\begin{center}
\epsfbox{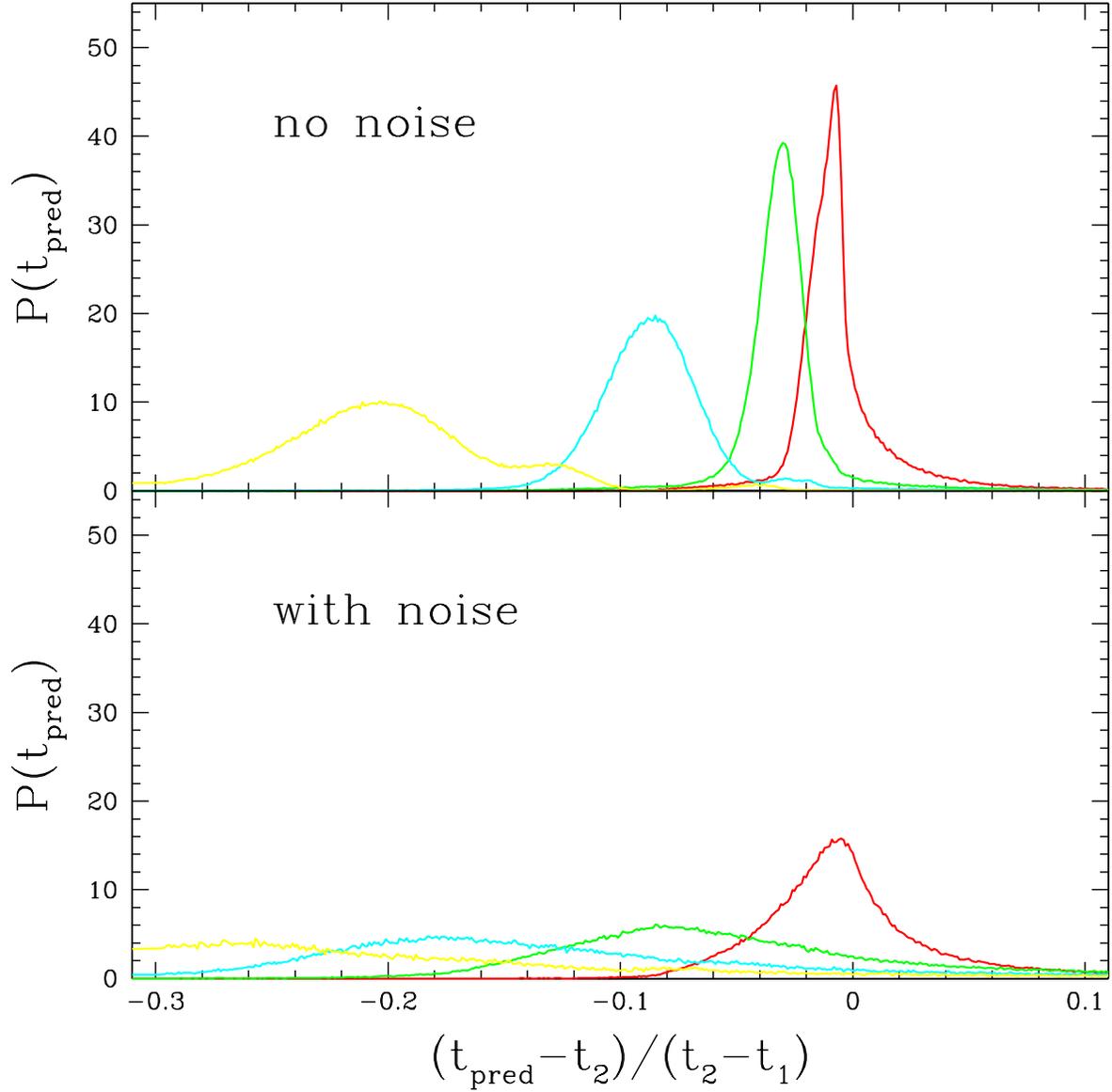}
\end{center}
\caption{Simulations of fits using approximate formula (eq.~\ref{eq:sqrt})
to predict 
the time of the second caustic crossing. The upper panel corresponds 
to simulations neglecting the observational errors. In the simulations 
presented in the lower panel the Gaussian scatter in measured stellar 
magnitudes $\Delta m =0.04$ is assumed. Each curve shows distribution of
predicted time of crossing based on observations made in limited span of
time during source brightening. The predictions shift systematically
from the left ("early") to the right ("late") if done later. On average
the predictions give too early crossing, so they are "safe". (See text
for details.)}\label{fig:sqrt}
\end{figure}

We use models with broad distribution of separations ($d \in (0.3,6.)$)
and mass ratios ($q \in (0.1, 1.0)$). For each model we draw at random
$\approx 500$ source paths crossing caustics, with different directions
and encounter parameters. For each path we find $s_1$ and $s_2$ - the
positions of crossings along the source trajectory, and $s_{12}$ - at
the magnification minimum. We use sources substantially (20 to 100
times) smaller than the distance between caustic crossings - 
otherwise the ``U-shape'' of the light curve would not be very pronounced.
We divide the path between $s_{12}$ and $s_2$ into six equal sections 
and draw randomly  ``points of observations'' from each of them, calculating
also the corresponding magnifications. For any three points belonging to
three consecutive sections we fit our formula and find the predicted 
time of the crossing $t_{\rm pred}$. The distributions of predictions 
made this way are shown in the upper panel of Fig.~\ref{fig:sqrt}.
 Different curves correspond to fits based on points drawn from different
fragments of the source trajectory. The
predictions become more accurate and less scattered, when one uses 
observations closer to the crossing, as expected.  All the
predictions are systematically shifted toward ``too early''.
In the lower
panel we show the results of a similar procedure but each ``observation''
has now ``measurement error'' included. (Errors in stellar magnitudes are
modeled as Gaussian with $\sigma_m=0.04$, typical for microlensing
observations.) The observational noise introduces
an extra scatter to the predictions, which also become more biased.
Applying the method to a larger number of observation points would
effectively diminish the noise and give results intermediate between
those shown in the upper and lower panels.

We apply the light curve fit based on eq.~(\ref{eq:sqrt}) to real data for 
BUL-38 and BUL-46. Again, as in the case of fitting binary lens models,
we choose various data subsamples each containing $\approx 10^1$ points,
this time using only the observations obtained when the source was 
brightening toward the second caustic crossing. For these particular events
the sensible fits are possible only late, at most two nights before the
crossing, and only the predictions on the last day are correct. 
(In the case of BUL-38 the fit based on data terminating two
days before the crossing gives a too early prediction, probably due to the
large error in the single point of the night $JD=2451737$. In the case of
BUL-46 the data terminating on $JD=1751$ leads to a fit with very broad
$\chi^2$ minimum, allowing for crossing between $JD=2451752$ and
$2451760$.) One can see again, that correct prediction of the second
caustic crossing can only be made very late.

\section{DISCUSSIONS}

We have tried to fit binary lens models to the light curves representing
caustic crossing events using incomplete data. Our purpose is to check
whether it is possible to predict the night of the second caustic
crossing and if this is the case, how early the reasonable prediction
can be made. The answer we obtained is not very promising: in the two
cases considered the unambiguous predictions could only be made based on
the data including the observations from the last night before the
crossing. These ``late'' predictions are true in the sense that they agree
with  the complete data sets for the events considered. The predictions
made earlier are unreliable since they give unacceptably large spread in
the time of the second crossing. 

The simplified approach using the square root formula is much less time
consuming but is also unable to give a reasonable early prediction of 
the second crossing of the caustic. Our simulations of such approach are
too simplified: the application to the real data gives worse
predictions than expected, probably due to the uneven time distribution
of observations.

The example of 2000-BUL-38 shows that the uneven coverage of the light
curve, especially the situation when the last observation used in
fitting is separated from the earlier data may strongly bias the
predictions.  The extrapolation is always strongly dependent on the last
known data point, so it is important to have this point measured
accurately. Having more than one data point per night certainly improves
the situation. On the other hand the models one gets from the fitting
procedure depend on the particular data sets and their errors, so the
predictions biased either way are probably unavoidable.

The fact that the binary lens models are not well constrained is 
known from the papers devoted to the subject (see in particular
Dominik 1999; Albrow et al. 1999b). The  models giving acceptable
fits to the given light curves belong to the quite large regions in the
$(q,d)$ plane. Our study shows that the regions of acceptable models can
have either broad or narrow spread in the second caustic crossing time, 
depending on the chosen subsample of data used in the fit. On the other
hand it is well known that the full coverage of the crossing is
sufficient to obtain a precise estimate of its time, regardless of the
lens model. Our estimates seem to converge to the right solution, but
slowly. 

In our approach we have neglected completely the possible motion of the
binary system. The inclusion of the binary rotation may improve the fit
since it offers more parameters to the model (e.g. Afonso et al.
200). Since our models are weakly constrained, at least for the data
samples considered, we do not expect extra parameters to improve the
situation. Similarly, without well sampled caustic crossing we do not
attempt to fit the limb darkening parameter. We use an ad hoc method for
the initial guess of the source size. We allow for the limited changes
in this parameter during the refinement of our models. Since the
refinement procedure is applied only to a limited number of candidate
models chosen on the basis of the approximate $\chi^2$ value, and the
allowed variation of the parameters on each step is very limited, we can
not claim that our models are optimized for the source size to the same
extent as for other parameters. Comparing the source radii of our best
models for different data sets with the radii obtained with the fits to
the second caustic crossings, we see the agreement between them up to a
factor two. 

Our simulations show, that the early predictions of the expected time for
the second caustic crossing are not possible. The predictions become
reliable only very shortly before the caustic crossing. The safest
observational strategy is then to sample as
densely as possible once a binary light curve starts to rise from the
inter-caustic minimum.

\section*{Acknowledgments}

We thank Bohdan Paczy\'nski for encouragements and 
helpful discussions. This work would not be possible
without the data released in real-time by the OGLE collaboration.
This work was supported in part by the NSF grant AST 98-20314 
and the Polish KBN grant 2-P03D-01316. 

{}

\bsp

\label{lastpage}
\end{document}